\documentclass[aps,amssymb,amsmath,twocolumn,pra,superscriptaddress]{revtex4}
\usepackage{graphicx}
\usepackage{xcolor}
\usepackage{mathtools}
\usepackage{epsfig}
\usepackage{dcolumn}
\usepackage[up]{subfigure}
\usepackage{float}
\usepackage{dsfont}	
\usepackage{wrapfig}
\usepackage{relsize}
\usepackage{epstopdf}
\usepackage{float}
\usepackage{ragged2e}
\usepackage{braket}
\usepackage{setspace}

\usepackage{hyperref}
\hypersetup{
    colorlinks=true,
    linkcolor=blue,
    citecolor=blue,
    urlcolor=blue
}

\makeatletter

\begin{document}


\title{Observation of room temperature gate tunable quantum confinement effect in photodoped junctionless MOSFET}

\author{Biswajit Khan$^ \dag $}
\affiliation{Centre for Applied Research in Electronics, Indian Institute of Technology Delhi, New Delhi 110016, India.}
\altaffiliation{B.K. and A.M. contributed equally to this paper.}
\author{Abir Mukherjee$^ \dag $}
\affiliation{Centre for Applied Research in Electronics, Indian Institute of Technology Delhi, New Delhi 110016, India.}

\author{Yordan M. Georgiev}
\affiliation{Institute of Ion Beam Physics and Materials Research, Helmholtz-Zentrum Dresden-Rossendorf(HZDR), Dresden 01328, Germany}

\author{Jean-Pierre Colinge}
\affiliation{Tyndall National Institute, University College Cork, Lee Maltings, Dyke Parade, Western road, Cork T12R5CP, Ireland.}
\author{Suprovat Ghosh}
\affiliation{Centre for Applied Research in Electronics, Indian Institute of Technology Delhi, New Delhi 110016, India.}
\author{Samaresh Das$^{}$}
\email{samaresh.das@care.iitd.ac.in}
\affiliation{Centre for Applied Research in Electronics, Indian Institute of Technology Delhi, New Delhi 110016, India.}

\begin{abstract}
In pursuing room temperature quantum hardware, our study introduces the effect of tunable photodoping in gate-bias assisted quantum wire within a tri-gated architecture of n-type junctionless MOSFET. To comprehend this behavior, we developed a theoretical model based on nonequilibrium Green’s function formalism. Our findings demonstrate the transition of the device from semi-classical to quantum domain under gate influence, with optical occupancy of electronic sub-bands at room temperature. Tunable photodoping originates from varying resonant excitations due to reallocation of interacting states between valence and conduction band. Gate bias causes centralization of electrons, which experience reduced interfacial trouble, resulting in distinct current peaks for specific gate biases compared to dark conditions at various drain voltages. This study underscores the persistence of quantum confinement effects via semi-classical transport, even at room temperature.
\end{abstract}



\maketitle

\section{Introduction}
The quantum confinement effect (QCE) is an extraordinary phenomenon in solid-state devices at cryogenic temperatures \cite{elzerman2004single, xia1997quantum, petta2005coherent, angus2007gate, rustagi2007low, li2013low}. QCE manifests when the device dimensions are on the same order of magnitude as the De Broglie wavelength of the carriers (electrons or holes), resulting in the quantization of energy levels. At ultra-low temperatures, the energy gaps between these discrete levels are significantly more significant than thermal fluctuations, enabling the distinct identification of these quantized energy levels in experiments. However, at room temperature (RT), the thermal energy vastly exceeds the energy differences between these quantized energy levels, making it highly challenging to observe QCE. Nevertheless, achieving QCE at RT is crucial for advancing quantum electronics hardware. Fabricating such nanodevices with traditional junctions is a complex process, requiring precise doping profiles at sub-nanometer junction regions. In this context, junction-less transistors \cite{colinge2011junctionless, lee2009junctionless, amin2013junctionless, colinge2006low, das2016high} have gained significant attention due to their practical advantages. In the past few years, Silicon quantum dot (QD) with nanometer islands \cite{gorman2005charge, yang2016quantum} and junctionless nanowire (Si NW) \cite{nishiguchi2006room, shi2013coherent, shaji2008spin, macquarrie2020progress, schoenfield2017coherent, penthorn2019two, hu2007ge} based QDs show massive potential for applications in quantum computation \cite{elzerman2004single, petta2005coherent, veldhorst2015two}, quantum sensing \cite{degen2017quantum, elzerman2004single, gonzalez2016gate, berman1997single, zhang2005single}, and provide a platform for CMOS-compatible nanoelectronics \cite{nikonov2013overview}. QCE at RT in 5 nm P-silicon [110] NW on Silicon-on-insulator (SOI) substrates by top-down approach is demonstrated experimentally \cite{trivedi2011quantum, yi2011room}. \cite{buin2008significant, singh2008si} showed enhanced performance based on RT QCE-based Si NW on SOI. SOI-based n-Si(110) and n-Si(100) NW field-effect transistor (FET) showed RT single electron/hole transport \cite{suzuki2013experimental} utilizing RT QCE. When incident light interacts with the phototransistor, photogenerated electrons and holes disperse into the channel region by the band profile, thus modifying the carrier distribution. This, in turn, leads to a modulation in conductivity, a phenomenon known as photogating \cite{konstantatos2012hybrid, jiang2022enhanced, yuan2018high, huang2016highly, guo2016black}.
\begin{figure}[H]
    \centering
    \includegraphics[scale=0.3]{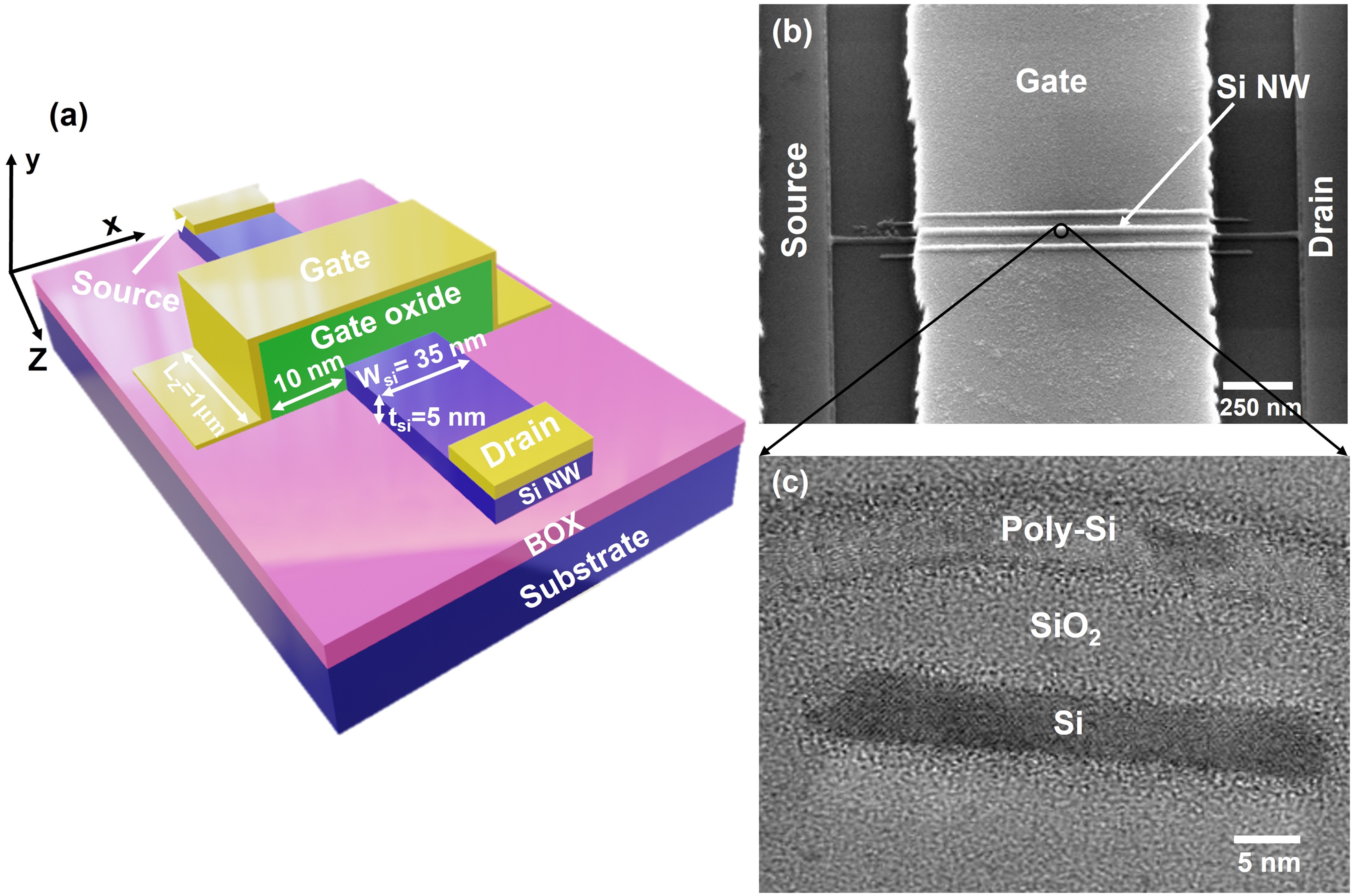}
    \caption{Device structure and electron microscopic images.} (a) Schematic of the device structure, (b) Typical scanning electron microscopic (SEM) image of the device, (c) Cross-sectional Transmission electron microscope (TEM) image of the device.
    \label{fig1}
\end{figure}

In transistors featuring silicon nanowires, the $Si/SiO_{2}$ interface provides a self-assembled electric field due to band bending at the interface, and this field accumulates photogenerated charges depending on the band bending; these are called photoinduced confined charges. The gate bias at the Si/SiO2 interface effectively facilitates one type of photogenerated carrier, while the opposite type accumulates at the center of the nanowire. As a result, this gate bias tunable potential well at the interface acts as a recombination barrier between the photo-generated carriers. Consequently, in other terminology, these confined charges act as a photogate voltage, altering the carrier profile and, ultimately, the channel's conductance. In addition to the confined charge at the interface, there are some interface trap charges that may contribute to the photogate voltage. In our case, these trap charges are significantly low\cite{jang2011low}. Introducing one type of carrier into the channel region via a photogate voltage, a process termed photo doping of the channel. Photoinduced doping is an enhanced form of doping in materials such as semiconductors under the influence of light \cite{aftab2022programmable, wu2017multi, yan2009nanowire, calarco2005size}. With the existing CMOS-compatible tri-gated MOSFET junctionless MOSFET fabrication procedure, here we have explored unique quantum phenomena under optical bias. Our experiment is premised on harnessing the response characteristics of a low-dimensional material that constitutes the gate in a junction-less transistor with the influence of impinged light. The illumination of the $p^+$-type polysilicon gate on a $n^+$ silicon nanowire leads to a deviation of the I-V characteristics of the system from the unilluminated case.
\begin{figure*}[t]
    \includegraphics[scale=0.6]{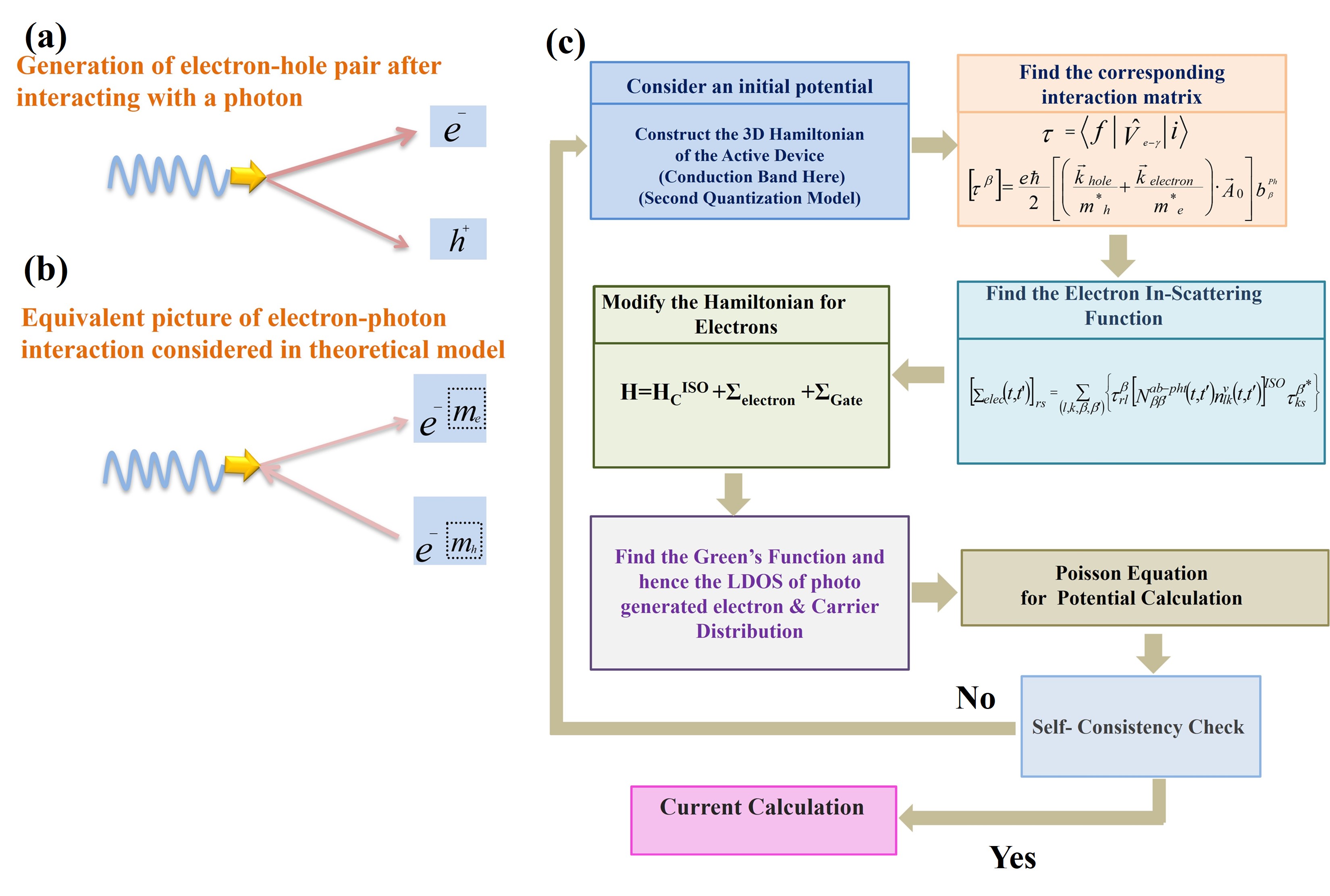}
    \caption{Interaction picture used in our model and the flow of the NEGF program.
        (a) The conventional picture of electron-hole pair generation due to light.
        (b) Feynman diagram corresponding to equivalent interaction picture utilized for theoretical modeling.
        (c) Total theoretical modeling and Schrödinger-Poisson self-consistency for quantum mechanically modulated semi-classical diffusive transport.}
    \label{fig2}  
\end{figure*}

\section{THEORETICAL BACKGROUNDS}

To understand the light-matter interaction, we need to find the corresponding interaction matrix, which has been conceptualized as a scattering picture as follows: 
\begin{equation}
    \tau = \langle f | \hat{V}_{e-\gamma} | i \rangle
     \label{eq:equation-1}
\end{equation}

The state preceding the scattering event depicts the virtual electron possessing the effective mass of the hole and momentum that is equal but opposite., i.e., $\ket{i}=\ket{-\vec{k}^{hole}_{i},m^{\star}_{h}}$ and the final state (post-scattering) of the real electron, $\ket{f}=\ket{\vec{k}^{electron}_{f},m^{\star}_{e}}$ as shown by Feynman diagram in figure-\ref{fig2}a.

The carrier-photon interaction operator is:  $\hat{V}_{{e-\gamma}}$ =$\frac{e}{2m^{\star}}(\hat{\Vec{p}}.\Vec{A}+\Vec{A}.\hat{\Vec{p}})$\\
Here, $\hat{\Vec{p}}$  is the momentum operator, and $\vec{A}$ is the incident photons' vector potential.

Thus, the light-matter interaction matrix can be written as \cite{aeberhard2012nonequilibrium, bertazzi2020nonequilibrium, kolek2019implementation}.
\begin{equation}
    [\tau^{\beta}]=\frac{e\hbar}{2}[(\frac{\Vec{k}_{hole}}{m^{\star}_{h}}+\frac{\Vec{k}_{electron}}{m^{\star}_{e}}).\Vec{A}_{O}]b^{ph}_{\beta}
    \tag{2} \label{eq:equation-2}
\end{equation}

The motion of charges in the quantum system is described by the Schrödinger-Poisson equations in the following manner:
\begin{align}
    \hat{H}\ket{\psi_{n}} & = E_{n}\ket{\psi_{n}} \tag{3} \label{eq:equation-3} \\
    \partial_{l}\epsilon_{lm}\partial_{m}\Phi(x_{k}) & =-( \pm e\rho(x_{k})) \tag{4} \label{eq:equation-4}
\end{align}

\begin{equation}
     H^{(C/V)}_{iso}=-\frac{\hbar^2}{2m_{0}}\sum_{l,m}(\partial_{l}{(m{^*}_{lm})^{-1}}\partial_{m})+\phi^{(C/V)}(x,y) 
      \tag{5} \label{eq:equation-5}
\end{equation}where $x_{k}\equiv(x_{1},x_{2},x_{3})$ in Cartesian coordinates. $\pm$ depicts the charge polarity of the hole and electron, respectively. The complete Hamiltonian for this system can be expressed in the following manner :
\begin{equation}
\begin{split}
H = & \sum_{i} H^{c}_{iso} c^{\dag}_{i} c_{i} + \sum_{j} H^{v}_{iso} v^{\dag}_{j} v_{j} + \sum_{\beta} \hbar \omega_{\beta} b^{\dag}_{\beta} b_{\beta} \\
& + \sum_{k} H^{G}_{iso} c^{\dag}_{k} c_{k} + \sum_{i,j,\beta} (\tau^{\beta}_{ij} c^{\dag}_{i} v_{j} b_{\beta} + \tau^{\beta\star}_{ji} c^{\dag}_{i} v_{j} b^{\dag}_{\beta} \\
& + \tau^{\beta}_{ij} v^{\dag}_{j} c_{i} b_{\beta} + \tau^{\beta\star}_{ji} v^{\dag}_{j} c_{i} b^{\dag}_{\beta}) 
\end{split}
\tag{6} \label{eq:equation-6}
\end{equation} where $(l,m)$ represent $(x,y,z)$. $\phi^{(C/V)}(x,y)$ is the distribution of potential energy of the conduction/valance band in $x$ and $y$ directions. $\epsilon_{lm}$ and $(m^*)_{lm}$ are the permittivity tensor and effective mass tensor, respectively, whose components vary at interfaces (e.g., semiconductor-insulator and metal-insulator). The possible supply of electrons and holes in a definite state ‘i ’ in the conduction band and state ‘j’ in the valence band are now, firstly from the state itself, secondly by absorption or emission of photons, and thirdly from the reservoirs.
\begin{equation}
\begin{split}
i\hbar\frac{d}{dt}c_{i} & = H^{C}_{iso}c_{i} + \sum_{j,\beta}\left(\tau^{\beta}_{ij}v_{j}b_{\beta} + \tau^{\beta\star}_{ij}v_{j}b^{\dag}_{\beta}\right) + \sum_{r}\zeta_{ik}c_{k}\\
i\hbar\frac{d}{dt}v_{j} & = H^{V}_{iso}v_{j} + \sum_{j,\beta}\left(\tau^{\beta}_{ij}c_{i}b_{\beta} + \tau^{\beta\star}_{ij}c_{i}b^{\dag}_{\beta}\right) + \sum_{r}\zeta_{jk}c_{k}
\end{split}
\tag{7} \label{eq:equation-7}
\end{equation}
The absorptions/emissions of photons in a definite mode are added up with those associated with annihilation of electrons from a definite state and subsequent creation of another. The situation is
expressed as,
\begin{equation}
i\hbar\frac{d}{dt}b_{\beta} = E_{\beta}b_{\beta} + \sum_{i,j}\left(\tau^{\beta\star}_{ij}c^{\dag}_{i}v_{j} + \tau^{\beta\star}_{ji}v^{\dag}_{j}c_{i}\right)
\tag{8} \label{eq:equation-8}
\end{equation}
The equation of motion for the electrons in the reservoirs is given by,
\begin{equation}
i\hbar\frac{d}{dt}c_{k}  = H^{G}_{R}c_{k} + \sum_{i}\zeta^{\star}_{ki}c_{i}+\sum_{j}\zeta^{\star}_{kj}v_{j}
\tag{9} \label{eq:equation-9}
\end{equation}
Here laser source (optical density) and valence band (occupied hole density) jointly carry the role of acting reservoir via joint density of states.
Here, $c_{i}$, $v_{j}$, and $c_{k}$ denote electron annihilation operators corresponding to the $i^{th}$ state of the conduction band, $j^{th}$ state of the valence band of the fin body, and $k^{th}$ state of the reservoirs (gate), respectively. Additionally, $b_{\beta}$ is the annihilation operator for photons at the $\beta$ mode. The term $\zeta_{ik}$ represents the interaction between reservoirs and the active device (conduction band). It is worth noting that the operator's b's, c's, and v's adhere to the Bose-Einstein (BE) commutation and Fermi-Dirac (FD) anti-commutation relations,
$[b_{\beta}, b^+_{\beta'}]= \delta_{\beta \beta'},\{c_i, c^+_{i'}\}=\delta_{ii'},\{v_j, v^+_{j'}\}= \delta_{jj'}$. The Keldysh formalism is now employed to evaluate the two-time correlation functions \cite{keldysh1965diagram}.
\begin{equation}
 \begin{aligned} 
&n^{c}_{im}(t,t^{'})&=&\langle c^{\dag}_{m}(t^{'}) c_{i}(t)\rangle,n^{v}_{jn}(t,t^{'})=\langle v^{\dag}_{n}(t^{'}) v_{j}(t)\rangle\\
&N^{ab\text{-}pht}_{\beta\beta^{'}}(t,t^{'})&=&\langle b^{ph\dag}_{\beta^{'}}(t^{'}) b^{ph}_{\beta}(t)\rangle 
\end{aligned}
\tag{10} \label{eq:equation-10}
\end{equation} 
$n^{c}_{im}$, $n^{v}_{jn}$, and $N^{ab-pht}_{\beta \beta^{'}}$ denote two-time correlation functions corresponding to the occupied state of electrons in the conduction band and valence band, as well as the absorbed photon, respectively. 
The electron and hole-in-scattering functions can be written as below\cite{aeberhard2008microscopic,mera2016hypergeometric,aeberhard2012nonequilibrium,bertazzi2020nonequilibrium,kolek2019implementation,sikdar2021design,sikdar2017analytical,ghosh2024optically}
\begin{equation}
\begin{split}
[\Sigma_{\text{elec}}(t,t')]_{rs} & =\sum_{(l,k,\beta,\beta')}\left\{ \tau^{\beta}_{rl} \left[ N^{ab\text{-pht}}_{\beta \beta'}(t,t')n^{v}_{lk}(t,t') \right]_{\text{iso}} \tau^{\beta'\star}_{ks} \right\} \\
[\Sigma_{\text{hole}}(t,t')]_{pq} & =\sum_{(l',k',\beta,\beta')} \left\{ \tau^{\beta}_{pl'} \left[ N^{ab\text{-pht}}_{\beta \beta'}(t,t')n^{c}_{l'k'}(t,t') \right]_{\text{iso}} \tau^{\beta'\star}_{k'q} \right\}
\end{split}
\tag{11} \label{eq:equation-11}
\end{equation}

Where the number of absorbed photons is given as:
$N^{ab\text{-}pht}_{\beta \beta^{\prime}} = \frac{\biggl\{\left(\frac{I_{\beta}}{E_{\beta}}\right) V_{ab} \delta_{\beta \beta^{\prime}}\biggr\}}{\tilde{v}}$ $I_{\beta}$, $n^{c}$, $n^{v}$, and $V_{ab}$ are the intensity of the laser source, correlation function of the filled states in the conduction band, valence band, and the absorbing volume, respectively. At the same time, $\tau$ represents the interaction potential governing the device's photo generation (electron-hole pair). The velocity of light within the material is $\tilde{v} = \frac{c}{n^{\lambda}}$, where $n^{\lambda}$ is the refractive index of Si for the corresponding wavelength of the laser source. After the addition of reservoirs and in scattering function for electrons and holes, the modified green’s function can be written as\cite{datta2000nanoscale,datta2005quantum}
\begin{equation}
\begin{split}
G^{M}_{elec}(E)=([E_{elec}I]-[{H_{C}}^{iso}]-[\Sigma_{elec}(E)]-[\Sigma^{R}(E)]])^{-1}\\
G^{M}_{hole}(E)=([E_{hole}I]-[{H_{V}}^{iso}]-[\Sigma_{hole}(E)]-[\Sigma^{R}(E)]])^{-1} 
\end{split}
\tag{12} \label{eq:equation-12}
\end{equation}
Hence, the concentrations of photo-generated electron and hole are calculated as follows\cite{sikdar2021design, ghosh2024optically}
\begin{equation}
\begin{split}
p^{ph}=[{\int_{-\infty}^{E_{V}}}{[G^{M}_{hole}(E)][\Sigma_{hole}(E)][G^{M}_{hole}(E)]^{\dag}} dE]\\
n^{ph}=[{\int_{E_{C}}^{+\infty}}{[G^{M}_{elec}(E)][\Sigma_{elec}(E)][G^{M}_{elec}(E)]^{\dag}}dE] 
\end{split}
\tag{13} \label{eq:equation-13}
\end{equation}
The transverse electronic state, n(x,y) can be split up into two transverse subspace as follows\cite{sikdar2019understanding}
\begin{equation}
    |n^{\tau}\rangle = |n_{x}\rangle \otimes |n_{y}\rangle
    \tag{14} \label{eq:equation-14}
\end{equation}
The 2D carrier distribution in the fin body cross-section (i.e., x-y plane) is given by,
\begin{equation}
n_{2D}(x,y)=\sum_{n_{\tau}}f(E_{c}+E_{n_{\tau}}-E_{f})\langle{{n_{\tau}}}|{x,y\rangle}{\langle{x,y}|{n_{\tau}}\rangle}
\tag{15} \label{eq:equation-15}
\end{equation}
Then the charge concentration $(1/cm^{3})$ $\rho^s = \frac{qn_{2D}}{L_{z}}$ at the source side and it was calculated followed by equation-\ref{eq:equation-13} and incorporated into equation-\ref{eq:equation-2} to obtain the electrostatic potential and then compared with the previous potential in the iteration loop to find the non-trivial difference self consistently, which is shown in figure-\ref{fig2}c. Considering the x and y principal crystallographic directions (e.g.,\(\langle 100 \rangle\), \(\langle 010 \rangle\), and \(\langle 001 \rangle\)), the effective mass tensor has only non-zero diagonal values. Incorporating that into the effective mass-based Hamiltonian and solving it using Green’s function method, the individual positional Fock space was determined and coupled to form the entire mesh to define the transistor in the x-y plane through carrier localization.

Hence, the total core-electron density at the source side can be written as: 

\begin{equation}
Q_{S} = \frac{q}{L_z}\int_{h_1}^{h_2} \int_{w_1}^{w_2} n_{2D}(x,y) \, dx \, dy
\tag{16} \label{eq:equation-16}
\end{equation}
where, $w_{2}-w_{1}=w_{eff}$ and $h_{2}-h_{1}=h_{eff}$ define the effective width and height respectively. $w_{2}$, $w_{1}$, $h_{2}$ and $h_{1}$ were found by numerically finding the non-zero carrier locations in both x and y- directions.

Therefore, the charge concentration at the drain side for long-channel devices can be calculated by \cite{trevisoli2012surface}
\begin{equation}
\begin{split}
Q_{D}=Q_{S}-(V_{FB}-V_{G}+\Phi_{s})C_{ox}
 \end{split}
 \tag{17} \label{eq:equation-17}
\end{equation}

Therefore, the surface potential-based drain current analytical expression is given by\cite{trevisoli2012surface},
\begin{equation}
    I_{D}={\frac{\mu}{L_{z}}}\biggr[{\frac{Q^{2}_S-Q^{2}_D}{2C_{ox}}}\biggl]
    \tag{18} \label{eq:equation-18}
\end{equation}

\begin{figure*}[t]
    \centering
    \includegraphics[scale=0.76]{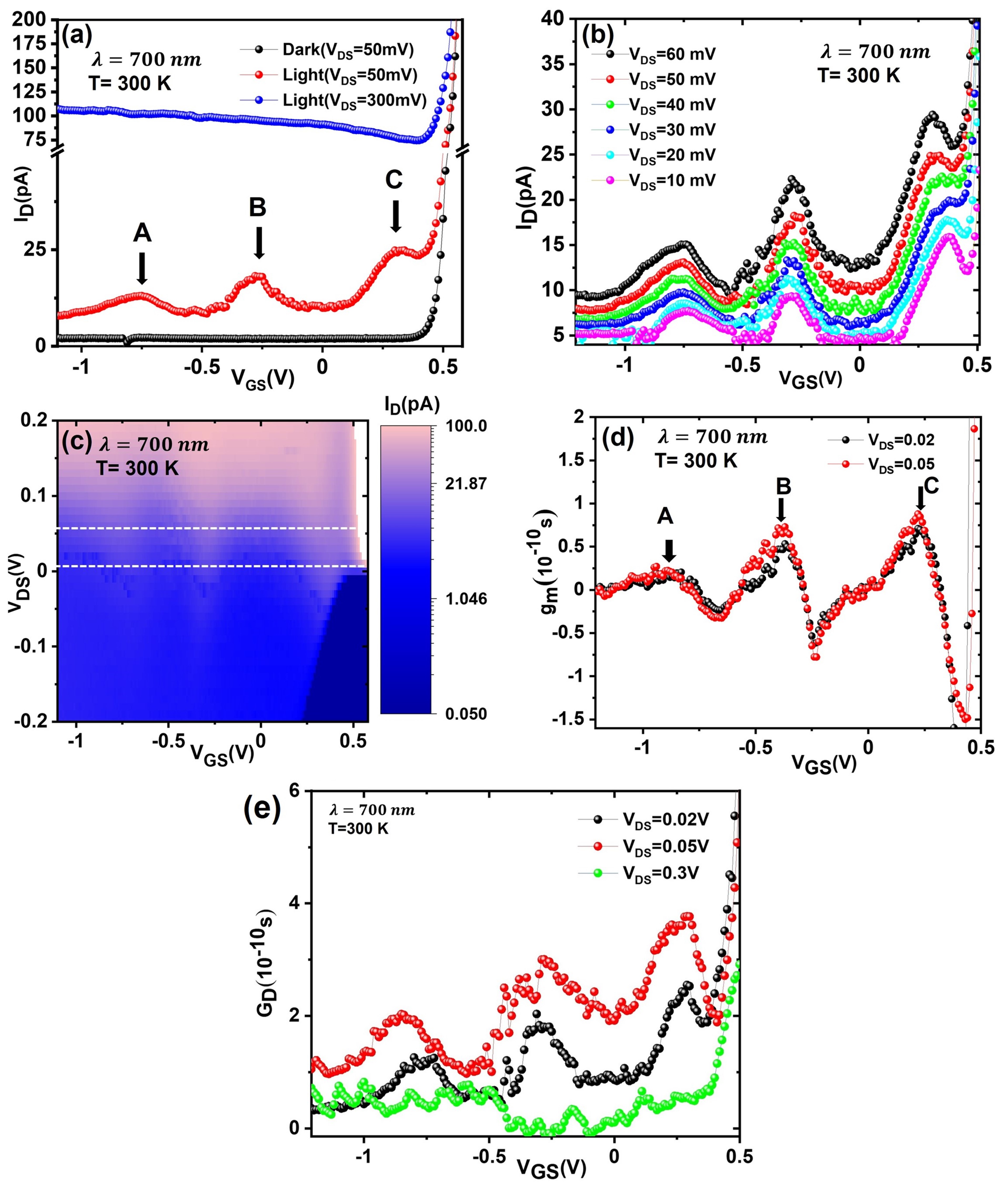}
    \caption{Photocurrent measurements.(a) Transfer characteristics under dark (black) and illumination (red) conditions at  $V_{DS}=50 mV$ and illuminated condition(blue) at $V_{DS}=300 mV$.(b) Transfer characteristics under illumination conditions for $V_{DS}= +10 mV$ to $60 mV$  (c) Contour plot: Drain current in the bias plane of drain voltage and gate voltage (d) Transconductance vs gate voltage plot at $V_{DS}= +10 mV$(black) and +$50 mV$(red) respectively. (e) Conductance vs gate voltage plot at $V_{DS}= +20 mV$ (black), + 50 mV (red) and $V_{DS}= +300 mV$ (green) respectively.}
    \label{fig3}
\end{figure*}

\begin{figure*}[t]
    \centering
    \includegraphics[scale=0.5]{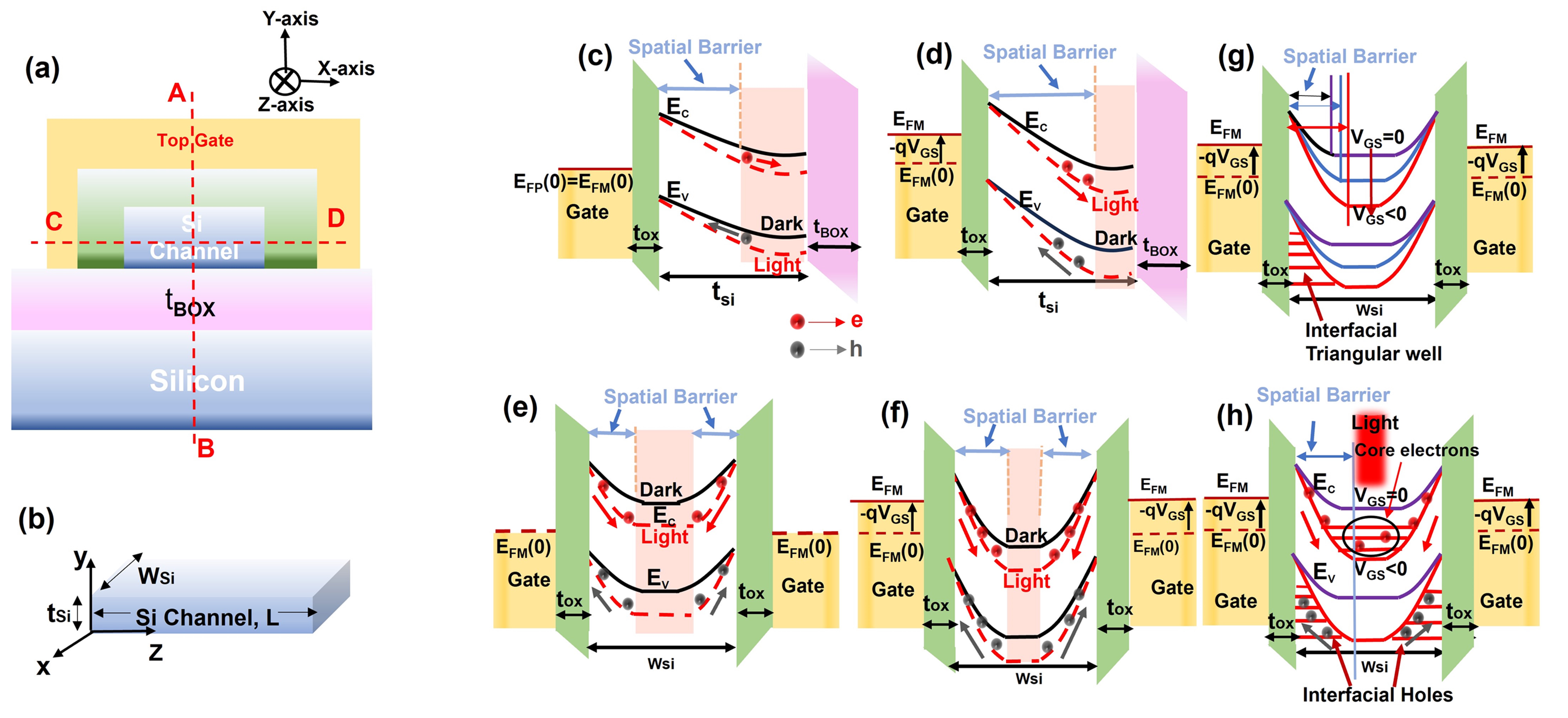}
    \caption{Cross-sectional schematic and energy band diagrams.(a) Cross-sectional schematic of the device, where AB cut line is along $t_{si}$(y-axis), and CD cut line is along $W_{si}$(x-axis) (b) Schematic of the Si channel with characteristic directions:$t_{si}$(y-axis),$W._{si}$(x-axis) and channel, $L$(z-axis). The asymmetry in the band diagram is due to the single gate configuration along this cut line AB. $E_{FP}$(0)=$E_{FM}$(0) is Fermi level of $P^{+}$ polysilicon gate when the gate voltage is zero. The solid black line is for dark, and the dashed red line is for the illuminated conditions, respectively, when (c) Gate voltage $V_{GS}$=0 and (d) Gate voltage $V_{GS}$ is negative. The symmetry in the band diagram is due to the double gate configuration along this cut-line CD. Energy band diagram along the cut line CD: The solid black line is for dark, and the dashed red line is for the illuminated conditions, respectively, when (e) Gate voltage $V_{GS}$=0 and (f) Gate voltage $V_{GS}$ is negative. (g) Energy band diagram along $W_{si}$(x-axis) from $V_{GS}$=0 to $V_{GS}<0$, showing the corresponding modification of triangular well states, hence modification of height and width of spatial barrier and changes in the slope of the energy band diagram near the $Si/Sio_{2}$ interface, which gives rise to a voltage-dependent interfacial triangular well. (h) Energy band diagram along $W_{si}$(x-axis) from $V_{GS}$=0 to $V_{GS}<0$, showing the corresponding modification of parabolic well for core electrons in the presence of light.}
    \label{fig4}
\end{figure*}

\begin{figure*}[t]
    \centering
    \includegraphics[scale=0.72]{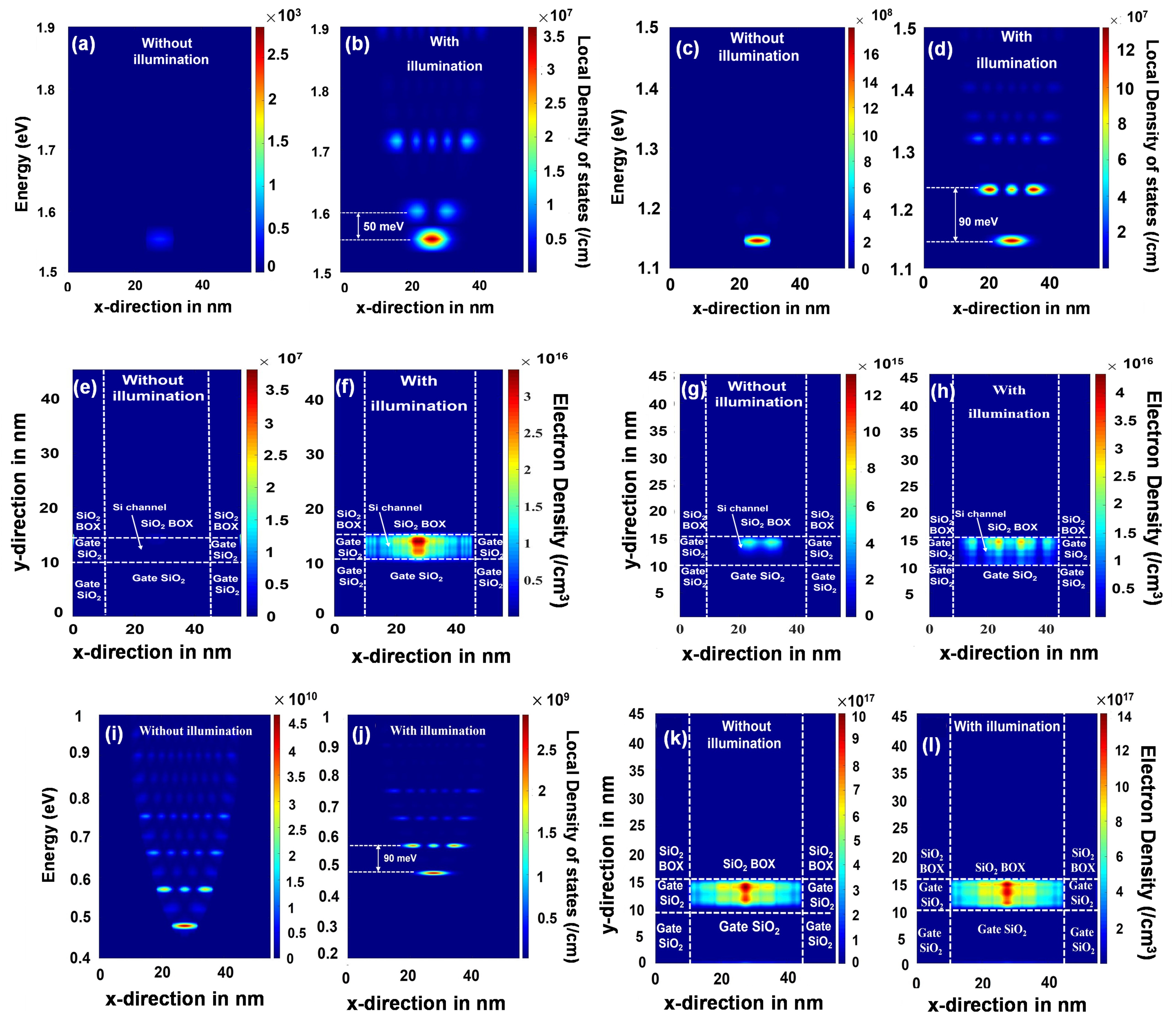}
    \caption{LDOS and carrier density under dark and light conditions from our developed NEGF Model. Local density of state (LDOS) along $W_{si}$: By incorporating the eigenvalues from $t_{si}$(y-space) and L(z-space), we reached an effective 3D Hamiltonian to get this coupled LDOS, (a) LDOS at dark condition (b) Photo LDOS(PLDOS) at illumination condition at $V_{GS}=-0.64V$. (c) LDOS at dark condition (d) Photo LDOS(PLDOS) at illumination condition at $V_{GS}=-0.26V$. Electron density contour plot in the cross-section plane of the nanowire(in XY plane or in $t_{si} \times W_{si}$ plane)(e) Without illumination (f) With the illumination of 700 nm laser at $V_{GS}=-0.64$V and (g) Without illumination (h) With the illumination of 700 nm laser at $V_{GS}=-0.26$V respectively. (i) LDOS at dark condition (j) Photo LDOS(PLDOS) at illumination condition at $V_{GS}=0.28V$. Electron density contour plot in the cross-section plane of the nanowire(in XY plane or in $t_{si} \times W_{si}$ plane)(k) Without illumination (l) With the illumination at $V_{GS}=0.28V$.}
    \label{fig5}
\end{figure*}

\section{RESULTS AND DISCUSSION}

\subsection{\textbf{\large{Device architecture and photo measurements:}}}
In this study, we investigate the behavior of room-temperature Quantum Confinement Effect (QCE)-based transport in the presence of light. As illustrated in figure \ref{fig1}a, we present a schematic representation of the tri-gated junctionless transistor we fabricated. The dimensions of the silicon nanowire (NW) are \text{$5 \, \mathrm{nm} \times 35 \, \mathrm{nm}$} (\(t_{\text{si}} \times W_{\text{si}}\)), and the gate length is \(1 \, \mu\mathrm{m}\). In contrast, figure \ref{fig1}b displays a scanning electron microscopic (SEM) image of the actual device, and Figure \ref{fig1}c is a cross-sectional transmission electron microscope (TEM) image of the device. 

Figure \ref{fig3} presents the results of photocurrent measurements for our junctionless transistor under different drain biases, specifically, at 50 mV and 300 mV when exposed to 700 nm light at room temperature (RT). Notably, a significant increase in current is observed under illumination compared to the dark current. Of particular interest is the presence of oscillations in the drain current concerning variations in gate voltage under illumination. To further investigate the influence of gate voltage on the drain current over a range of drain biases, we systematically varied the drain bias from -200 mV to 200 mV and recorded the transfer characteristics. This comprehensive examination is illustrated in the plane of drain and gate biases, as shown in figure \ref{fig3}c.

The graph's white lines delineate the drain bias values range where we observed significant gate tunability. Figure \ref{fig3}b reveals substantial gate tunability in the photocurrent from \(V_{\text{DS}}=+10 \, \mathrm{mV}\) to \(V_{\text{DS}}=+60 \, \mathrm{mV}\) at RT. For higher drain bias, this tunability vanished \cite{colinge2006low, colinge2007quantum, colinge2006room, rustagi2007low, lee2010dual, trivedi2011quantum, ma2015observation, li2013low, je2000silicon}. To see those peaks more prominently, we plotted the transconductance with respect to gate voltage for drain voltage 20 mV and 50 mV, respectively, in figure \ref{fig3}d. In addition, channel conductance behavior concerning gate voltage is illustrated in figure \ref{fig3}e for drain voltage 20 mV,50 mV, and 300 mV, respectively.

\subsection{\textbf{\large{Photo doping in the core of nanowire.}}}  
Compared to bulk silicon, nanostructured silicon (e.g., nanowires, quantum dots) exhibits significant absorbance for 700 nm light and shorter wavelengths\cite{das2016high,tsakalakos2007strong,hasan2013review,sikdar2017analytical}. For such ultra-small structures of silicon, both reflection and transmission (i.e., high absorption) are reduced to nearly zero \cite{tsakalakos2007strong}. Additionally, the average absorption of light with a wavelength range of 300–700 nm by a 5 nm pitch of silicon nanowire is 27.95 \%. \cite{fan2018optimal}. Furthermore, Dhyani et al.\cite{dhyani2019diameter} showed that in nanowires with a smaller diameter, the incoming photon energy contained within the volume is noticeably greater. A 700 nm laser was chosen for two reasons: it generates photocarriers much above the band gap for visualizing electronic sub-band filling via voltage-dependent separation and ensures silicon has adequate absorptance with less thermal noise interference, enabling room-temperature observation of quantum effects. In Dhyani et al.\cite{dhyani2019diameter}, it is also demonstrated that 700 nm has a peak spectral response for lower-dimensional nanowires. Therefore, in the presence of laser illumination, an electron-hole pair (e-h pair) will be generated, and the electrons will accumulate in the core of the channel, where holes reach a triangular potential well near the interface, as shown pictorially in figure-\ref{fig4}c to \ref{fig4} h. To understand this electron accumulation at the center of NW more clearly, we illustrated the energy band diagrams along the cutline, denoted as the red dotted lines, namely AB and CD, respectively, in figure-\ref{fig4}a. As the photo-generated hole moves toward the surface due to upward band bending (negative surface potential), consequently, the photo-generated electrons move toward the core of the nanowire (figure-\ref{fig4}c to \ref{fig4}f). Therefore, photo-generated holes accumulate at the gate bias-dependent well at the Si/Sio2 interface while photo-generated electrons move towards the core of NW. Due to this spatial quarantine, which is depicted as a spatial barrier in figure-4(c-h)) of photo-generated e-h pairs, an additional positive gate voltage is induced by accumulated holes at the interface of $Si/Sio_{2}$ of the NW channel, which in turn modulates the concentration of electrons at the core of n channel NW. Excess electrons due to light do the job of doping at the core of the nanowire (NW), termed photodoping. When the negative gate voltage is applied to the top gate, the slope of the energy band near the interface will be increased (figures-\ref{fig4}g, \ref{fig4}h, \ref{fig4}d and \ref{fig4}f) compared to the no gate voltage case (figure-\ref{fig4}c and \ref{fig4}e) to support the high electric field. Due to this, the availability of states near the interface will increase (figures-\ref{fig4} g and \ref{fig4} h). Therefore, for a high electric field, more photo-generated holes will be translated toward the $Si/Sio_{2}$ interface and facilitated into the increased states at the interface. Consequently, more electrons will be accumulated at the center of the channel (figure \ref{fig4}h) due to the increase in efficiency in e-h pair separation and the rise in positive photo gate voltage. This electron accumulation at the center of NW is illustrated by the light shadow of pink color in figure-\ref{fig4}d and \ref{fig4}f. Negative gate voltage modifies the parabolic potential well by increasing the depletion width and introducing a strong spatial barrier between the holes at the interface and the electron at the core. In other words, band bending near the interface will increase as we increase the negative gate voltage, introducing a high recombination barrier for photo-generated e-h pairs. Therefore, the efficiency of separating the photo-generated e-h pair will increase, and eventually, the concentration of electrons at the center will increase, too.
\begin{figure*}[t]
    \centering
    \includegraphics[scale=0.5]{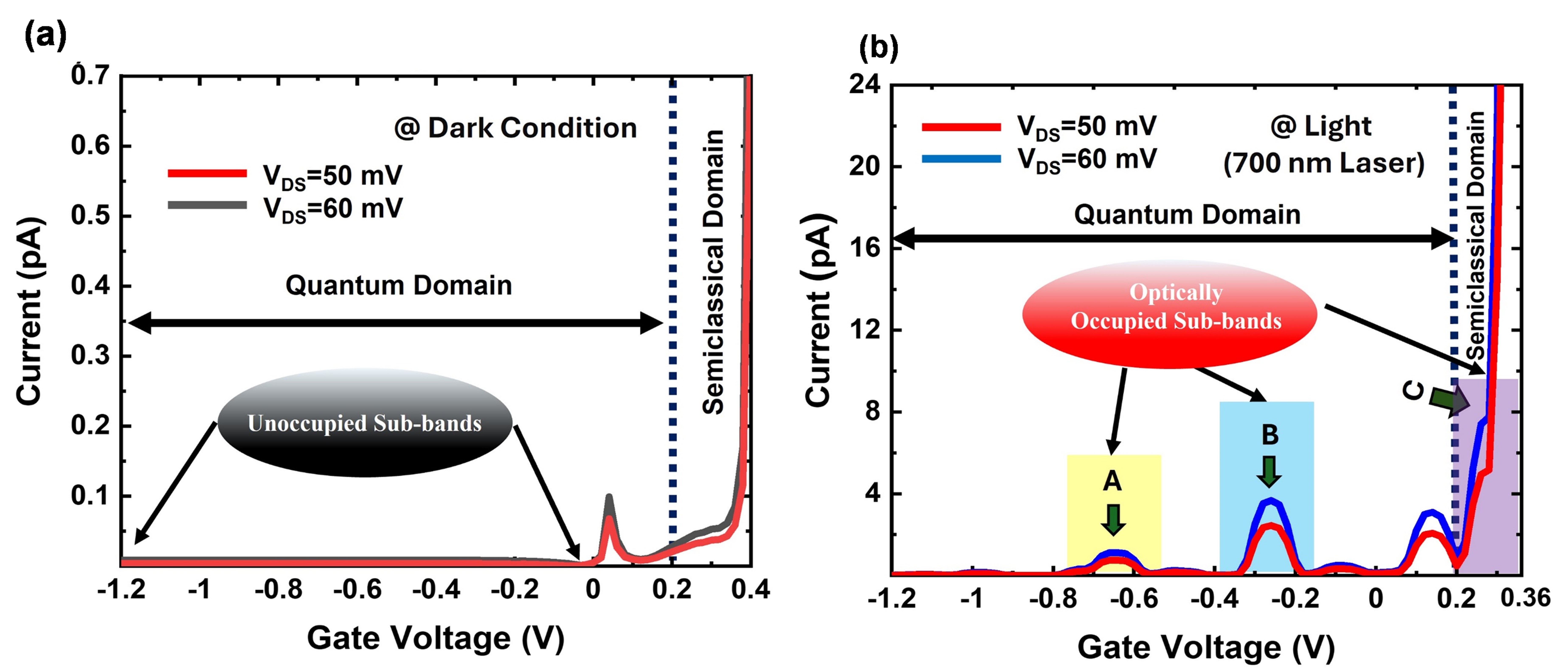}
    \caption{Current under dark and light conditions from our theoretical model.Calculated current from the charge density based on our theoretical model (a) without illumination (b) With illumination at $V_{DS}=50 mV$ and $60 mV$, respectively.}
    \label{fig6}
\end{figure*}

\subsection{\textbf{\large{Occupancy of subbands in the presence of light:}}}

A channel of photo-generated electrons is induced at the center of the nanowire due to photo-doping effects resulting from modification of the potential well of NW due to gate voltage. This occurs through the bulk conduction mechanism of the junctionless transistor(JLT) \cite{colinge2011junctionless,colinge2010nanowire,trevisoli2012surface,rai2022comparative,colinge1990conduction,colinge2012silicon,gnani2011theory}. In the nanoscale field-effect devices, the motion of electrons is guided by quantum conditions as well as the electrostatic fields simultaneously. For this purpose an initial assumption of arbitrary potential variation along the channel with fixed boundary conditions applied at the reservoirs is needed. The equation is solved by non-equilibrium Green’s function (NEGF) formalism to find the local density of electronic states (LDOS) within the channel at various sub-bands transported from the reservoirs. Thus, carrier density corresponding to the assumed potential across the channel is calculated from the LDOS and the respective fermionic distribution of the reservoirs. The carrier density, then, is put into Poisson’s equation to solve for the electrostatic potential along the channel under the same boundary conditions and compared with the assumed potential to verify the consistency. If the difference between the assumed potential values and the calculated ones are non-negligible, then the latter is put into the Hamiltonian and the same process is run until the input and output values of potential become consistent. In the nano-structured semiconductor devices charge transport behavior cannot be explained solely by electrostatic interactions. Poisson’s equation describes the interaction between a number of charges, which shows that, more the number of charges, more is the potential energy. However, quantum properties restrict the electrons to stay at same state and constrain by the condition that, more the potential energy less is the probability of the particle to stay. The inner contradiction between this two-fold property of the charged fermions (i.e. electrons and holes), that is to stay (since electrostatic condition) and not to stay (since quantum condition), is the origin of their motion. This indeed is considered in the present model to predict the light matter interaction and transport behavior that includes the quantum and electrostatic interactions together and needs to achieve their self-consistency to find the final results regarding the device characteristics. Our modeling starts constructing a closed system Hamiltonian, and then we build the light-matter interaction term, namely, the electron/hole in-scattering function. This function accounts for the excess electrons/holes entering the system due to interaction. Ultimately, we add this to the closed system Hamiltonian to make it an open system. After that, we perform Schrödinger-Poison's self-consistency to precisely evaluate the local density of states (LDOS) and carrier distribution in dark and illumination conditions. Figure-\ref{fig5} shows such LDOS and carrier density under dark and light conditions using our developed NEGF model. By incorporating the eigenvalues from $t_{si}$(y-space) and L (z-space), we achieved an effective 3D Hamiltonian to obtain coupled LDOS along $W_{si}$. This figure shows the coupled local density of states (LDOS) along $W_{si}$ and carrier concentration under dark conditions and photo LDOS (PLDOS) under illumination conditions at $V_{GS}=-0.64V and -0.26V$ respectively.
 
\subsection{\textbf{\large{Device Current from model:}}}
Using equations-\ref{eq:equation-15} to \ref{eq:equation-18}, we have calculated $I_{D}$ and plotted it against the gate voltage $V_{GS}$ for $V_{DS}$=50 mV for nonilluminated case (figure-\ref{fig6}a) and illuminated case (figure-\ref{fig6}b) respectively. From those two figures, we can see that $I_{D}$ for dark is almost flat due to the deficiency of electrons and the unavailability of states. This is illustrated in figure-\ref{fig5}a and \ref{fig5}e for peak A, corresponding to the $V_{GS}$=-0.64V. It is also further illustrated for $V_{GS}$=-0.26V corresponding to peaks B, in figure-\ref{fig5}c and figure-\ref{fig5}g, respectively. Likewise, figure-\ref{fig5}i and \ref{fig5}j 
 show corresponding LDOS and figure-\ref{fig5}k and \ref{fig5}l show corresponding electron density contour plot in the cross-section plane at dark and illumination condition respectively for peak C.

 $I_{D}$ from our theoretical calculations for illuminated case shows nearly exact peaks, namely A, B, and C, respectively, for $V_{DS}$=50 mV (Figure-\ref{fig6}b). Here, the Gate voltage is pushing the device to the quantum regime along $W_{si}$ direction, and being already in the quantum domain in $t_{si}(=5 nm)$ direction, we finally reach voltage tunable quantum wire. A coupled mode LDOS of that quantum wire from our NEGF model is illustrated in figure-\ref{fig5}a and figure-\ref{fig5}b for the non-illuminated and illuminated case respectively for gate Voltage $V_{GS}$=-0.64V. This illustration demonstrates the energy difference between the initial and subsequent sub-bands is 50 meV for gate Voltage $V_{GS}$=-0.64V. Similarly, the energy spacing of the first and third sub-band is 90 meV for gate Voltage $V_{GS}$=-0.26V, which is shown in figure-\ref{fig5}c and figure-\ref{fig5}d respectively. The energy spacing of the first and third sub-band is 90 meV for gate Voltage $V_{GS}$=0.28V, which is illustrated in figure-\ref{fig5}j. These peaks A, B, and C are related optically populated subbands for different gate voltage $V_{GS}$=-0.64V,-0.26V, and 0.28V, respectively, at room temperature.

\section{Conclusions}
In summary, we conducted an experimental study on gate-tunable semi-classical transport in a silicon nanowire top-gated junctionless n-channel MOSFET, focusing on the quantum confinement effect when exposed to light at room temperature. Notably, such oscillations in current persisted under small drain-to-source bias but became less discernible at higher values. Additionally, we developed a theoretical model for light-electron interactions within our nanostructure device and its corresponding transport characteristics. The gate in our device produces an electric field that causes a reorganization of interacting states, resulting in Quantum occupancy effect (QOE) that can be manipulated by voltage, even at room temperature. By applying negative gate bias, our structure converts from a 1D-confined to a 2D confined structure, thereby achieving a voltage-tunable quantum wire. Hence, we focused mainly on the negative gate-bias region (i.e., fully depleted condition), where sub-band occupation is possible only via optical perturbation. The origin of such drain current oscillation arises from such optical occupancy event at both x and y directions, where sub-band filling along x-direction experiences better tunability due to finite interaction of parabolic potential well of electrons at the core and triangular potential wells of holes at the interfaces. Therefore, this letter unveils an abnormality in quantum mechanically controlled diffusive transport that persists even at room temperature under the influence of illumination. This will open up new possibilities for quantum devices at room temperature by photo-doping in low-dimensional field-effect transistors. Nevertheless, the transparent gate may achieve equivalent photodoping with significantly reduced optical power. This can be used for sophisticated quantum photonic applications in extremely low lighting conditions.

\subsection{\textbf{\large{Acknowledgement:}}}

The authors thank the Defence Research and Development Organisation (DRDO) and the Department of Science and Technology (DST) for financial support. BK acknowledges the financial assistance from the Prime Minister Research Fellows (PMRF) Scheme (PMRF ID: 1401633), India. Additionally, special appreciation is expressed to James Haigh from Hitachi Cambridge Laboratory for assisting with optical measurements.\vspace{8pt}.

\textbf{\Large{Competing interests.} }\\
The authors declare no competing interests.\vspace{8pt}

\textbf{\Large{Data Avaibility.} }\\
The datasets produced in this research can be acquired by making a reasonable request to the corresponding author.

\textbf{Correspondence} and inquiries for materials should be directed to S.D.

\section*{\textbf{APPENDIX A: Device Fabrication}}
Figure-\ref{fig1}a  shows the schematic of a tri-gated nanowire junctionless transistor. The nanowire is highly doped ($\sim 3\times10^{19}\,\mathrm{cm}^{-3}$) such that we can have a high amount of ON current. The gate material surrounded the nanowire from three directions, as shown in figure-\ref{fig1}, to increase the gate control over the channel, and this tri-gate has the same functionality as a conventional single gate. Therefore, this tri-gate geometry can easily control the flow of carriers from source to drain with the highest precision. We use $n$-type nanowire or channel and p-type gate in this work. Therefore, we now put a $p$-type gate on the $n$-type channel; due to work function difference and very low nanowire thickness (5 nm), the channel beneath the gate got depleted completely. So, this type of junction-less device works in accumulation mode, which is unlikely like other devices with junctions. To fabricate tri-gated silicon nanowire-based junctionless MOSFET, we used silicon-on-insulator (SOI) wafers with a few high-quality nanometers of top silicon. We did ion implantation of arsenic to dope the top silicon of SOI wafer into n-type silicon, and resultant arsenic doping was in the order of $3\times10^{19}$ $cm^{-3}$. After making the top silicon into n-type, we did nanolithography using electron beam lithography (EBL) to define the nanowire, where we spin-coated top silicon with negative electron resist Hydrogen Silsesquioxane (HSQ). Next, we did reactive ion etching (RIE) of EBL patterned nanowire, and finally, silicon nanowire (Si NW) was realized. After realizing the nanowire, we did dry oxidation of Sio2 on the top of the nanowire to get 10 nm top gate oxide. Then, we patterned the oxide over the nanowire. In the next step to realizing 1 um top gate, we deposited 50 nm amorphous silicon in a low-pressure chemical vapor chamber at $550^{\circ}c $. We doped the amorphous silicon heavily with boron to make $P^{+}$ top gate. After this, we annealed the sample for 30 mins at $900^{\circ}c$ in nitrogen to make it polycrystalline silicon from amorphous silicon. Then, we did EBL and RIE to pattern the polysilicon to define the 1-um gate. To isolate the gate from source and drain contact, we deposited a very slim layer of sio2. To make source and drain contact, we opened a window in the sio2 via photolithography and deposited Ti/W-Al. The scanning electron microscopy (SEM) image of fabricated tri-gated silicon nanowire MOSFET is shown in figure-\ref{fig1}b.

\section*{\textbf{APPENDIX B: Photo Response Measurement}}
The photo responsiveness of the Tri-Gated Junctionless (JL) nanowire n-channel transistor we fabricated was assessed under light exposure with a wavelength of 700 nm. Illuminating our Nanowire MOSFET with a laser beam with incident power on the nanowire is 4.12 nW. This illumination induces electron excitation, resulting in the production of a measurable current. Subsequently, the source measuring unit captures and examines this current, offering valuable information about NW MOSFET’s photoconductivity and electronic characteristics. In Figure 2, the transfer characteristics of our fabricated device are depicted under constant illumination conditions, while maintaining a fixed drain-to-source voltage of 50 mV. Even at incident power on the nanowire is 4.12 nW, our device exhibited a discernible photo response compared to the non-illuminated case, highlighting its high sensitivity.
To further understand the photo response, we varied the drain-to-source voltage from -200 mV to +200 mV, maintaining the same illumination conditions ( laser beam with incident power on the nanowire is 4.12 nW and a wavelength of 700 nm). Determining the efficiency of light absorption by a material involves considering two key factors: the absorption coefficient of the material and the distance traveled by light within the material. This relationship can be expressed through a simple equation 
\begin{equation}
    P_{channel}=P_0(1-e^{\alpha {t_d}})
   \tag{19} \label{eq:equation-19}
\end{equation}
Where $P_0$ is the incident power on the surface of the channel material,$\alpha $ is the absorption coefficient of the channel material, $t_d$ is the thickness of the silicon nanowire channel, and $P_{channel}$ is the effective power absorbed by a channel region. Now, as the incident power is in the nanowatt (nW) region, therefore the effective power absorbed by the nanowire is much less than the incident power according to the incident power law (equation-19).\\

\bibliography{BK_Main}

\end{document}